\documentclass{iau} 
\usepackage{graphicx}
\usepackage{url}
\usepackage{xcolor}

\title[~~The Deeper Wider Faster program] 
{The Deeper Wider Faster program:\\
chasing the fastest bursts 
in the Universe}

\author[Igor Andreoni \& Jeff Cooke]   
{Igor Andreoni$^{1,2,3}$
 \and Jeff Cooke$^{1,2,4}$}

\affiliation{$^1$Centre for Astrophysics and Supercomputing, Swinburne University of Technology, \\
PO Box 218, H29, Hawthorn, VIC 3122, Australia \\ 
email: {\tt igor.andreoni@gmail.com} \\[\affilskip]
$^2$Australian Astronomical Observatory, \\
105 Delhi Rd, North Ryde, NSW 2113, Australia \\[\affilskip]
$^3$The ARC Centre of Excellence for Gravitational Wave Discovery (OzGrav) \\[\affilskip]
$^4$The ARC Centre of Excellence for All-Sky Astrophysics (CAASTRO)
} 

\pubyear{2018}
\volume{339}  
\jname{Southern Horizons in Time-Domain Astronomy}
\editors{ }
\begin{document}

\maketitle

\begin{abstract}
We present the Deeper Wider Faster (DWF) program that coordinates more than 30 multi-wavelength and multi-messenger facilities worldwide and in space to detect and study fast transients (millisecond-to-hours duration).  DWF has four main components, (1)  simultaneous observations, where $\sim$ 10 major facilities, from radio to gamma-ray, are coordinated to perform deep, wide-field, fast-cadenced observations of the same field at the same time.  Radio telescopes search for fast radio bursts while optical imagers and high-energy instruments search for seconds-to-hours timescale transient events, (2) real-time (seconds to minutes) supercomputer data processing and candidate identification, along with real-time (minutes) human inspection of candidates using sophisticated visualisation technology, (3) rapid-response (minutes) follow-up spectroscopy and imaging and conventional ToO observations, and (4) long-term follow up with a global network of 1--4m-class telescopes.  The principal goals of DWF are to discover and study counterparts to fast radio bursts and gravitational wave events, along with millisecond-to-hour duration transients at all wavelengths.

\keywords{surveys, gravitational waves, supernovae: general, stars: variables: other, stars: flare}
\end{abstract}

\firstsection 
\section{Introduction}
Many interesting transient events occur at short timescales, ranging from hours to minutes, and even down to millisecond-timescales. These include for example supernova shock breakouts, gamma-ray bursts, flare stars, and collisions between type Ia supernovae and their companion stars. Recently a short gamma-ray bursts and its associated kilonova were demonstrated to be counterparts to gravitational waves (GWs, e.g. \cite[Abbott \etal\ 2017]{Abbott2017MMA}). Although it was observed that the r-processes generate an optical and infrared signal lasting for a few days, only a cadence of hours or minutes allows measurement and characterisation of the rising phase and the peak of the blue precursor\footnote{See the ``Open Kilonova Catalog'' for a complete light curve: \url{https://kilonova.space/}}.  In addition to what we learned from the study of the electromagnetic counterpart to the GW170817 event, a population of kilonovae and perhaps new types of counterparts to GW events still need to be uncovered with a systematic exploration in depth and timescale regime.  

Fast radio bursts (FRBs, \cite[Lorimer \etal\ 2007]{Lorimer2007}) constitute another example of short-timescale transient. FRBs are detected as dispersed radio flashes lasting only a few milliseconds. More than 30 FRBs have been discovered to date (\cite[Petroff \etal\ 2016]{Petroff2016}), but their nature is still unknown. Importantly, multi-wavelength and multi-messenger counterparts to FRBs and other fast transients may occur while (or even {\it before}) the burst is generated and/or detected, in which case simultaneous observations before, during, and after the radio burst occurs are necessary to study such phenomena.  

The discovery of the repeating FRB 121102 (\cite[Spitler \etal\ 2016]{Spitler2016}) and its association with a host galaxy marked an important step in understanding FRBs, confirming their extragalactic origin.  However, many questions remain unanswered, including the nature of the FRB event itself, which physical process generates the coherent radio emission, and whether all FRBs are ``repeaters'' or FRBs come in more than one population.

\begin{figure}[b]
\centering
\begin{center}
 \includegraphics[width=\columnwidth]{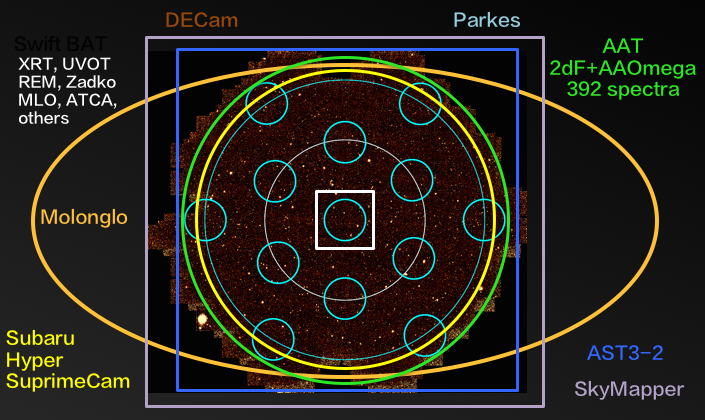} 
 \caption{The fields of view of the Parkes and Molonglo radio telescopes match very well with the fields of view of wide-field optical telescopes, such as the Blanco/DECam (image), Subaru/HSC, SkyMapper and AST3-2 telescopes shown here.  The {\it Swfit} BAT field of view is much larger than the the figure. The detection of a FRB during a DWF run with more than one radio telescope and several other multi-wavelength facilities targeting the same region of the sky with deep, fast-cadence observations, would help resolve the overall nature of FRBs in one shot.  DWF performs these coordinated observations and also processes the data and identifies candidates in minutes to trigger deep spectroscopic and imaging observations.}

\label{figure FoV}
\end{center}
\end{figure}


\section{The Deeper Wider Faster program}

The faint and rapidly-evolving time-domain regime has been poorly explored so far at most wavelengths, mainly due to the focus of past surveys on the science of supernova explosions and cosmology, along with technological limitations.  The Deeper, Wider, Faster program (DWF\footnote{\url{http://dwfprogram.altervista.org/}}, Cooke et al., in preparation) overcomes previous obstacles by coordinating multiple multi-wavelength world-class facilities and capitalising on new observational and computational resources and technology. 

The DWF coordinates more than 30 multi-wavelength and multi-messenger observatories to perform: 1) simultaneous, fast-cadence observations in time domain; 2) supercomputer data processing in real time (seconds to minutes) and human inspection (minutes) of the candidates using sophisticated visualisation technology; 3) rapid-response spectroscopic and photometric follow up and conventional target of opportunity (ToO) observations with radio, high-energy, and 1--10m optical, telescopes; 4) scheduled `interleaved' observations with nightly cadence, and long-term follow up to monitor the behaviour of transient and variable sources weeks after the DWF observing run.  The DWF observing campaigns usually occur for 4 to 6 consecutive nights per semester, which favors the discovery and characterisation of millisecond- to day-timescale transients and maximises the simultaneous coverage of target fields with highly competitive facilities at all wavelengths. The DWF program rapidly grew from 4 facilities in 2015 to more than 30 at the end of 2017.  We expect to coordinate more than 40 telescopes in 2018.  

\subsection{Simultaneous deep, multi-wavelength, fast-cadenced observations}

Figure\,\ref{figure FoV} shows the good match between the fields of view (FoV) of telescopes capable of real-time FRB detection and identification, such as the Parkes and Molonglo radio telescopes, world-class optical imagers Subaru/Hyper Suprime-Cam (HSC) and Blanco/Dark Energy Camera (DECam), along with other imaging and multi-object spectroscopy facilities at other wavelengths.  The involvement of key, major simultaneous observing facilities located worldwide has resulted in a division of DWF into the DWF--Pacific and DWF--Atlantic programs
in order to capitalise on the number, diversity, and geographical locations of the participating facilities.  These current programs include:  

{\bf DWF--Pacific simultaneous} -- {\bf radio:} Parkes, Molonglo, Australian Square Kilometre Array Pathfinder (ASKAP), Australia Telescope Compact Array (ATCA), Murchinson Widefield Array (MWA); -- {\bf optical/IR:} Subaru/HSC, Zadko telescope, Antarctica Schmidt Telescopes 2 (AST3-2), University of Tokyo 1m telescope, Mount Laguna Observatory (MLO), Anglo-Australian Telescope (AAT);
-- {\bf high energy:} {\it Hard X-ray Modulation Telescope} (HXMT) satellite, {\it Swift} satellite (BAT, XRT, and UVOT instruments), HAWC, and Pierre Augure Observatory.

{\bf DWF--Atlantic simultaneous observations} -- {\bf radio:} MeerKAT, VLA, (Green Bank); -- {\bf optical/IR:} Blanco/DECam, Rapid Eye Mount (REM) telescope, Astronomical Station Vidojevica (ASV), Antarctica Schmidt Telescopes 2 (AST3-2), Virgin Islands Robotic Telescope (VIRT), Gamma-ray Burst Optical/Near-infrared Detector (GROND); -- {\bf high energy:} {\it Hard X-ray Modulation Telescope} (HXMT) satellite, {\it Swift} satellite (BAT, XRT, and UVOT instruments), HAWC, and Pierre Auger Observatory.

Importantly, we coordinate the DWF campaigns with the LIGO/Virgo/GEO600 gravitational interferometers observing runs and other multi-messenger facilities, when possible.  This allows all facilities participating the DWF observations to join forces in a deep, multi-wavelength centrally coordinated follow up of GW triggers under the DWF memorandum of understanding (MoU) signed with the Ligo/Virgo Consortium.

\subsection{Real-time data processing and candidate identification}

The DWF program is an on-going project that underwent a rapid expansion and each of its components has constantly been improved.  The DWF core team made the coordinated observations described above possible and effective, while developing a dedicated infrastructure that includes: i) a ``lossy" compression code for rapid data transfer (\cite[Vohl \etal\ 2017]{Vohl2017}); ii) a pipeline for rapid transient identification in optical images (\cite[Andreoni \etal\ 2017a]{Andreoni2017mary}), iii) state-of-art visualization technology and design of new collaborative workspaces (\cite[Meade \etal\ 2017]{Meade2017}) to enable efficient human inspection of candidates.  In parallel, we manage dedicated projects to improve the identification of transient/variable sources with machine-learning techniques.  The web-based interface we are developing engages astronomers, students, and a broader community into the discovery and classification of transient and variable sources in real time and via archival analysis.  

\subsection{Follow up}

Key to the success of the project is rapid-response and long-term follow up of the discoveries. Some fast transients (e.g., the shock breakout at the birth of supernova explosions) are associated with longer-timescale events and therefore need to be monitored in the weeks following their identification. 

Currently, the following telescopes operate in one or more follow-up modes (rapid-response - minutes, conventional ToO - hours, `interleaved' - nightly, and long-term monitoring), depending on the DWF observing run:
Gemini-South, Keck, SALT, VLT, AAT, Palomar, Lick, Lijiang, ANU 2.3m, Xinglong, Gattini, HXMT, Swift, GROND, REM, ASV, SkyMapper, CNEOST, Zadko, LCOGT, UoT 1m, MLO, LCOGT, AST3-2, TNTS, VIRT, and Huntsman telescopes.  Others will be participating in the future.

\subsection{First results}

To date, no FRB has been detected in real time during DWF observations.  Nevertheless, we have unveiled thousands of transients and variables in the optical images acquired with DECam.  We have discovered dozens of fast optical transients evolving on seconds-to-minutes timescales, with nearly all sources being of Galactic origin (e.g., flare stars).  With the data in hand we can probe for the first time the minute-timescale time domain for extragalactic fast optical transients (Andreoni et al., in preparation).  Papers are currently in preparation to report the discovery of peculiar M-dwarf flares, early- and late-phase core-collapse and Type~Ia supernovae (among which those reported in ATel\#100072 and ATel\#100078) with multi-wavelength photometric and spectroscopic follow up, and several serendipitous discoveries.  Finally, 11 facilities involved in DWF contributed to the study of the electromagnetic counterpart to the GW170817 event (\cite[Andreoni \etal\ 2017b]{Andreoni2017gw}).

\section{Summary}

The DWF program coordinates $>30$ multi-wavelength, multi-messenger facilities around the world and in space to chase the fastest bursts in the Universe, including FRBs and counterparts to GW events.  We developed the necessary technology to enable coordinated observations with real-time analysis, rapid-response follow up, nightly interleaved and long-term follow up to catch and understand bursts on millisecond-to-hours timescales.  Over a hundred researchers from all continents have contributed to the DWF observing campaigns and have greatly helped the maturity and efficiency of the program.  DWF presents a new approach to time-domain astronomy that will shape how future transient detections and study will be performed.

\end{document}